\documentclass[aps,prd,twocolumn,noshowpacs,superscriptaddress,groupedaddress,nofootinbib,preprintnumbers]{revtex4}  
\pdfoutput=1
\usepackage{graphicx}  
\usepackage{dcolumn}   
\usepackage{bm}            
\usepackage{amsmath,amssymb,amsfonts,wasysym} 
\usepackage{mathrsfs}
\usepackage[bf,footnotesize]{caption2}
\usepackage{bbm}
\usepackage[]{hyperref}%
\hypersetup{
colorlinks=true,
linkcolor=red,
anchorcolor=black,
citecolor=blue,
filecolor=cyan,
menucolor=red,
runcolor=filecolor,
urlcolor=magenta,
bookmarks=true,
bookmarksnumbered=true
}
\def\l{\label}

\def\({\left(}
\def\){\right)}
\def\f{\frac}
\def\be{\begin{equation}}
\def\ee{\end{equation}}
\def\bry{\begin{array}}
\def\ery{\end{array}}
\def\bes{\begin{subequations}}
\def\ees{\end{subequations}}
\def\bit{\begin{itemize}}
\def\eit{\end{itemize}}
\def\ben{\begin{enumerate}}
\def\een{\end{enumerate}}
\def\dst{\displaystyle}

\def\ovl{\overline}

\newcommand{\Dsl}{D\llap{/\kern+1.5pt}}
\newcommand{\MET}{E\llap{/\kern1.5pt}_T}

\allowdisplaybreaks

\begin{document}


\preprint{IFT-UAM/CSIC-12-66}
\preprint{LPN12-075}

\vspace{2mm}
\title{Photophilic Higgs from sgoldstino mixing}

\author{Brando Bellazzini}
\email{brando.bellazzini@pd.infn.it}
\address{Dipartimento di Fisica e Astronomia, Universit\`a di Padova and INFN Sezione di Padova, Italy}
\address{SISSA, Via Bonomea 265, I-34136 Trieste, Italy}
\author{Christoffer Petersson}
\email{christoffer.petersson@csic.es}
\author{Riccardo Torre}
\email{riccardo.torre@csic.es}
\address{Instituto de F\'isica Te\'orica UAM/CSIC,
Universidad Aut\'onoma de Madrid, Cantoblanco, E-28049, Spain}

\date{\today}

\begin{abstract}
The spontaneous breaking of linearly realized $\mathcal{N}=1$ supersymmetry implies the existence of a pseudo-Goldstone fermion, the goldstino, and of its complex scalar superpartner, the sgoldstino. The latter has generically sizable tree-level couplings to Standard Model gauge bosons while its couplings to SM fermions are suppressed.
We consider a light sgoldstino, with a mass around 1 TeV, that mixes with a SM-like Higgs scalar at around 125 GeV. We show that such a mixing can enhance the Higgs to di-photon signal rate while evading all the relevant experimental bounds and without significantly affecting the other decay channels.
\end{abstract}

\maketitle

\section{Introduction}

The spontaneous breaking of $\mathcal{N}=1$ supersymmetry (SUSY) implies the presence of a pseudo-Goldstone fermion, the goldstino, which becomes the longitudinal component of the gravitino once SUSY is coupled to gravity. As a result of this super-Higgs mechanism, the gravitino acquires a mass (in flat space-time) \mbox{$m_{3/2}=f/(\sqrt{3}M_{\mathrm{P}})$}, where $\sqrt{f}$ is the SUSY breaking scale and $M_{\mathrm{P}}\approx 2.4\cdot 10^{18}$ GeV is the Planck mass. We consider the scenario where $\sqrt{f}$ is within one order of magnitude above the TeV scale. In such a case, due to the supersymmetric version of the equivalence theorem \cite{1977PhLB...70..461F,1988PhLB..215..313C}, the gravitino can be replaced by its goldstino components and SUSY can be treated as an approximate global symmetry. 

When SUSY is linearly realized, but spontaneously broken, the SUSY algebra implies the presence of the goldstino complex scalar superpartner: the sgoldstino. In contrast to the goldstino, this scalar is not protected by the Goldstone shift symmetry and therefore it generically acquires a mass of the order $f/M$ upon integrating out some heavy states, with a characteristic scale $M$, of the SUSY breaking sector. The precise coefficient depends on the details of this sector, such as the loop order at which the sgoldstino mass is generated. 
If this scale $M$ is comparable to the  mass scale of the heavy states that are integrated out in order to generate the soft masses of the Standard Model (SM) superpartners, then also the soft masses are of the order $f/M$. Again, the coefficients depend on, for example, whether the soft masses are generated at tree level \cite{Nardecchia:1207267} or loop level \cite{Giudice:1998fj}. Hence, whether the sgoldstino is heavier or lighter than some, or all, of the SM superpartners is a model-dependent question. Of course, if its mass is much larger than the energy scale under consideration, the sgoldstino can be integrated out and SUSY is non-linearly realized in the resulting low energy effective theory, see e.g.~Ref.~\cite{Komargodski:2009cq}. In contrast, if the sgoldstino is sufficiently light, it should be included in the low energy effective theory, see e.g.~Refs.~\cite{Brignole:1997hb,Brignole:2003hb,Bertolini:2011wj,Petersson:2011in} (see also Refs.~\cite{Perazzi:2000ku,Perazzi:2000dk,Gorbunov:2000ii,Gorbunov:2002co,Demidov:2004uy} for studies of sgoldstino collider phenomenology).

The couplings of the sgoldstino to the SM fields are determined by the supercurrent conservation. In particular, the strength of its couplings to SM gauge bosons is dictated by ratios of the gaugino masses over the SUSY breaking scale. The sgoldstino couples almost exclusively to the transverse polarizations of the gauge fields, implying that its couplings to gluons and photons are not 1-loop suppressed with respect to the couplings to the $Z$ and $W$ bosons. 

In this work we consider the case where the sgoldstino scalar is light, around the TeV scale, and mixes with a SM-like Higgs  boson (at around 125 GeV), modifying its phenomenology at the LHC. In particular, due to the smallness of the coupling of the SM-like Higgs boson to photons, even a small mixing with the sgoldstino can substantially enhance the Higgs to di-photon signal rate. In contrast, since the couplings of the SM-like Higgs to the $Z$ and $W$ bosons are large, the partial widths into ElectroWeak (EW) gauge bosons are not affected by such a small mixing. However, their signal rates would still be affected by an enhanced gluon fusion production cross section of the Higgs, arising again because of the Higgs-sgoldstino mixing. Such an overall enhancement in the production can be avoided by requiring at least one of either the bino or wino masses to be of the same order as the gluino mass.

A common way to enhance the Higgs to di-photon rate is by reducing the Higgs coupling to bottom and/or top quarks. In fact, a decrease in the rate for $h\rightarrow b\bar{b}$ simply reduces the overall Higgs rate and hence enhances all the subdominant Branching Ratios (BRs). On the other hand, a reduction in the top quark coupling, including a sign flip, increases the diphoton BR because of the interference of the top-loop with the W-loop that generate the effective Higgs coupling to photons (see e.g.~Refs.~\cite{Espinosa:2012tc,Azatov:2012us,Carmi:2012wc,Giardino:2012ux} for recent discussions concerning the general low-energy parametrization of the Higgs couplings).  In the context of the Next to MSSM (NMSSM), the possibility of enhancing the Higgs to di-photon rate by Higgs mixing was discussed in e.g.~Refs.~\cite{Ellwanger:2011sv,Gunion:2012ta}.  One can also increase the loop contribution to the Higgs coupling to photons by considering extra heavy states that are charged under electromagnetism and have a large coupling to the Higgs boson, see e.g.~Refs.~\cite{Bellazzini:2012tv,Carena:2012wl}. 
In this paper we show that the Higgs to di-photon enhancement can be achieved without significantly affecting any of the other decay channels, while satisfying experimental bounds on the sgoldstino decays, arising primarily from di-jet and di-photon searches.

Because of the general properties of the sgoldstino, we are not restricted to any particular SUSY extension of the SM, such as the MSSM or extensions thereof. However, we do require the presence of a SM-like Higgs scalar with a mass at around 125 GeV as suggested by the recent LHC results \cite{ATLASCollaboration:2012uh,CMSCollaboration:2012tb}. Note that, since the scale of SUSY breaking is low, tree level F-term contributions to the Higgs mass, arising from the goldstino superfield, can increase the usual MSSM tree level mass that arises from the D-term contributions of the vector superfields, see Refs.~\cite{Antoniadis:2010hs,Petersson:2011in,Petersson:2012tl}. 

The paper is organized as follows. In Section \ref{sec:sgold} we discuss SUSY operators which give rise to the sgoldstino couplings that control the most relevant decay channels.  We also determine the sgoldstino production cross section and compare it against the relevant experimental bounds. In Section \ref{mix}  we discuss the scenario where the sgoldstino mixes with a SM-like Higgs boson.  Motivated by the possible hint at the LHC for an excess in the Higgs to di-photon signal rate compared to the SM \cite{ATLASCollaboration:2012wx,CMSCollaboration:2012uz}, we study the possibility of enhancing this rate via the mixing with the sgoldstino. In Section \ref{Sec:Conclusion} we draw our conclusions and in the Appendix we provide the relevant sgoldstino couplings and decay widths.

\section{Sgoldstino phenonenology}\label{sec:sgold}

In this section we discuss the most relevant couplings, decay channels and the production mechanism of the sgoldstino. We consider the case in which the goldstino resides in a gauge singlet chiral superfield \mbox{$X=x+\sqrt{2}\theta G+\theta^2 F_X$}, where the auxiliary component $F_X$ acquires a non-vanishing VEV $f$. In contrast to the common way of parametrizing SUSY breaking using a background spurion, here we are treating both the goldstino $G$ and the sgoldstino $x$ as propagating degrees of freedom. The complex sgoldstino scalar gives rise to both a CP-even and a CP-odd\footnote{As a consequence of the structure of the sgoldstino vertices, the production and decays of the CP-odd state will be very similar to those of the CP-even state.} neutral scalar state in the mass basis. In the following we will restrict our analysis to the CP-even sgoldstino scalar $\phi$ only.

\subsection{Couplings and decays}

The most relevant sgoldstino couplings to the SM gauge fields are obtained by simply coupling the goldstino superfield to the gauge kinetic terms, 
\begin{eqnarray}
\label{susyop1}
 \sum_{i=1}^{3}\,-\frac{m_{i}}{2\,f}\int d^2\theta \, X   \,
W_{A_i}^\alpha W_{\alpha}^{A_i}+ \mathrm{h.c.} ~,
\end{eqnarray}
where the indices $A_1=1$, $A_2=1,2,3$ and $A_3=1,\cdots,8$ run over the adjoint representations of $U(1)_Y$, $SU(2)_L$ and $SU(3)_C$. By inserting the VEV of the auxiliary field $F_X=f$, one recovers the usual gaugino masses $m_{i}$. Since we are interested in the sgoldstino interactions we pick the scalar component of $X$ and the gauge field strength component of $W_{\alpha}^{A_i}$. All the relevant vertices are given in the Appendix, including the couplings between the CP-even sgoldstino and the longitudinal components of the $Z$ and $W$ bosons, i.e.~$\phi Z^{\mu} Z_{\mu}$ and $\phi W^{+\,\mu}W^{-}_{\mu}$, which arise after mixing with the SM-like Higgs and will therefore be suppressed by the mixing angle. Since the typical size of the mixing angle we will consider in the following is small, such couplings turn out to be negligible in our analysis. 

In order to account for the invisible decay of the sgoldstino into two goldstinos, we consider the following SUSY operator, which gives rise to a sgoldstino coupling to two goldstinos, 
\begin{equation}
\label{susyop2}
-\frac{m_\phi^2}{4f^2} \int d^4 \theta  \left( X^\dagger X \right)^2\, .
\end{equation}  
This operator also gives rise to a soft mass $m_{\phi}$ for the sgoldstino and is generically generated by integrating out heavy states in the SUSY breaking sector, e.g.~massive fields integrated out at 1-loop in O'Raifeartaigh models or a massive vector multiplet integrated out at tree level in models with an additional $U(1)$ gauge group. 

\begin{figure*}[!t]
\begin{center}
\includegraphics[scale=0.38]{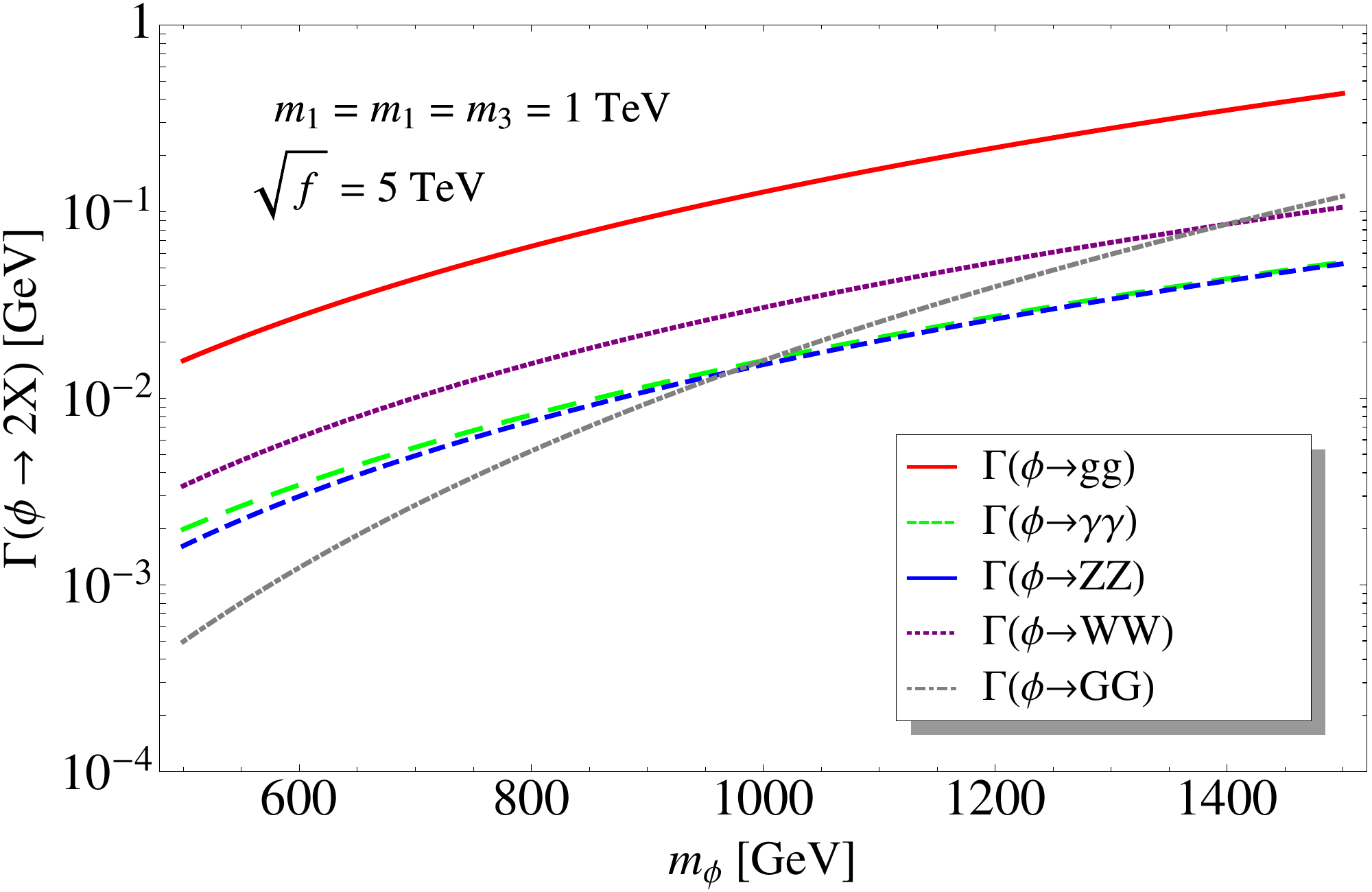}\hspace{1cm}
\includegraphics[scale=0.38]{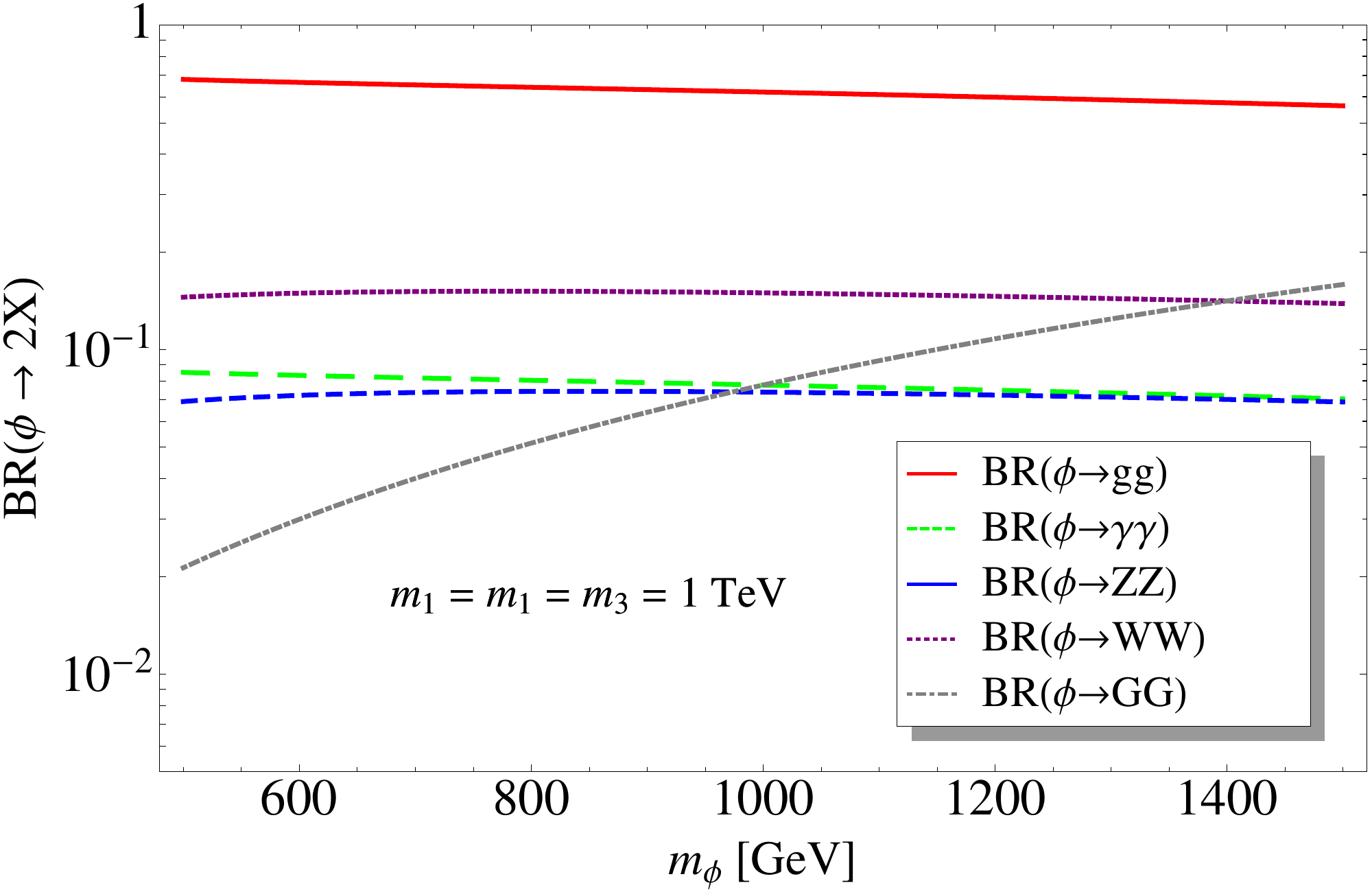}
\end{center}
\caption{ 
\small
In the left (right) plot, the most relevant partial decay widths (BRs) for the CP-even sgoldstino are given as a function of the sgoldstino mass. In both plots, all the gaugino masses have been set equal to 1 TeV and, in the left plot, $\sqrt{f}$ has been set equal to 5 TeV. 
}\label{fig:widthBRs}
\end{figure*}
The couplings of the sgoldstino to the SM femions arise from the following SUSY operators, which also provide the $A$-terms, 
\begin{eqnarray}
\label{susyop3}
-\frac{A_u}{f} \int d^2\theta  \,X \, Q \,H_u\, U^c +\mathrm{h.c.}~,
\end{eqnarray}
together with the analogous operators for the up-type quark and lepton chiral superfields. By picking the scalar component of $X$, the fermion components of the quark superfields and the VEV of the $h_u$-scalar, we see that all the sgoldstino couplings to SM fermions are suppressed by a factor $A\,v/f$. Hence, if the $A$-terms are proportional to the Yukawa couplings, it is only the top $A$-term which can be relevant. However, due to the additional $v/\sqrt{f}$ suppression, the sgoldstino coupling to top quarks turns out to be irrelevant in our analysis. Note that, via the usual Yukawa superpotential couplings, the sgoldstino can couple to the SM quarks by mixing with the SM-like Higgs. However, since we will assume that such a mixing is small, this will not induce any significant couplings. Hence, the sgoldstino will generically couple very weakly to fermions and therefore all sgoldstino decays into fermions will be subdominant. 

We see from Eqs.~\eqref{susyop1}, \eqref{susyop2} and \eqref{susyop3} that all the sgoldstino couplings are proportional to soft parameters over the SUSY breaking scale, which follows from the fact that the superpartner of the sgoldstino, the goldstino, couples to the divergence of the supercurrent. In order to have a valid perturbative expansion in terms of such ratios, we require all soft parameters to be smaller than $\sqrt{f}$. Also, in order to have the corrections to the kinetic terms under control, which arise from non-renormalizable SUSY operators, all VEVs of the scalars should be much smaller than $\sqrt{f}$. This latter requirement will not only be satisfied for the Higgs fields but also for the sgoldstino, which, from the analysis done e.g.~in Ref.~\cite{Petersson:2011in}, we expect in general to have a (exactly or at least approximately) vanishing VEV. 

In Figure \ref{fig:widthBRs}, the most relevant partial widths and BRs are given as functions of the sgoldstino mass in the mass range between 500 GeV and 1.5 TeV. The analytic expressions for these partial widths are given in the Appendix. In this figure we have chosen a particularly simple point in the parameter space, in which all the gaugino masses have been set to be 1 TeV \footnote{The choice of setting all the gaugino masses equal to each other will be motivated in the following section whereas the choice of 1 TeV is motivated by the fact that the LHC searches are constraining the gluinos to be above around 800 GeV, see e.g.~Ref.~\cite{Papucci:2011vz} for a rather model-independent analysis.} and we see that, due to the color factor enhancement of the gluon width, the decay into two gluons is generically the dominant decay channel. This parameter choice implies that it is simple to rescale the partial widths in the left figure for different values of $f$. Notice that the dependence on $f$ drops out in the BRs. Also note that the coupling governing the partial width of the sgoldstino decaying into $\gamma Z$ is  non-vanishing only for $m_1 \neq m_2$ (see Table \ref{Table1}) and therefore, due to the particular choice of parameters, this channel is not present in Figure \ref{fig:widthBRs}.

\subsection{Production and bounds}
Concerning the production mechanism for the sgoldstino, due to the coupling to gluons in Eq.~\eqref{susyop1}, the sgoldstino will be resonantly produced at hadron colliders. The Leading Order (LO) production cross section by gluon-gluon fusion can be written in the form\footnote{This formula makes use of the Narrow Width Approximation (NWA) which is a good approximation as long as $\Gamma_{\phi}\lesssim 0.1 m_{\phi}$ and $m_{\phi}\ll \sqrt{s}$.},
\be\l{prodCS}
\bry{lll}
\dst \sigma_{\phi}&=&\dst \f{\pi^{2}}{8}\f{\Gamma\(\phi\to gg\)}{s m_{\phi}}\\
&&\dst \hspace{-3mm}\times\int_{m_{\phi}^{2}/s}^{1}\f{dx}{x}f_{p/g}\(x,m_{\phi}^{2}\)f_{p,\bar{p}/g}\(\f{m_{\phi}^{2}}{xs},m_{\phi}^{2}\)\,,
\ery
\ee
where the partial width $\Gamma\(\phi\to gg\)$ is given by Eq.~\eqref{widthphigg}, $s$ is the center of mass energy squared and $f_{p/g}\(x,Q^{2}\)$ are the parton distribution functions defined at the scale $Q^{2}$.
\begin{figure*}[!t]
\begin{center}
\includegraphics[scale=0.38]{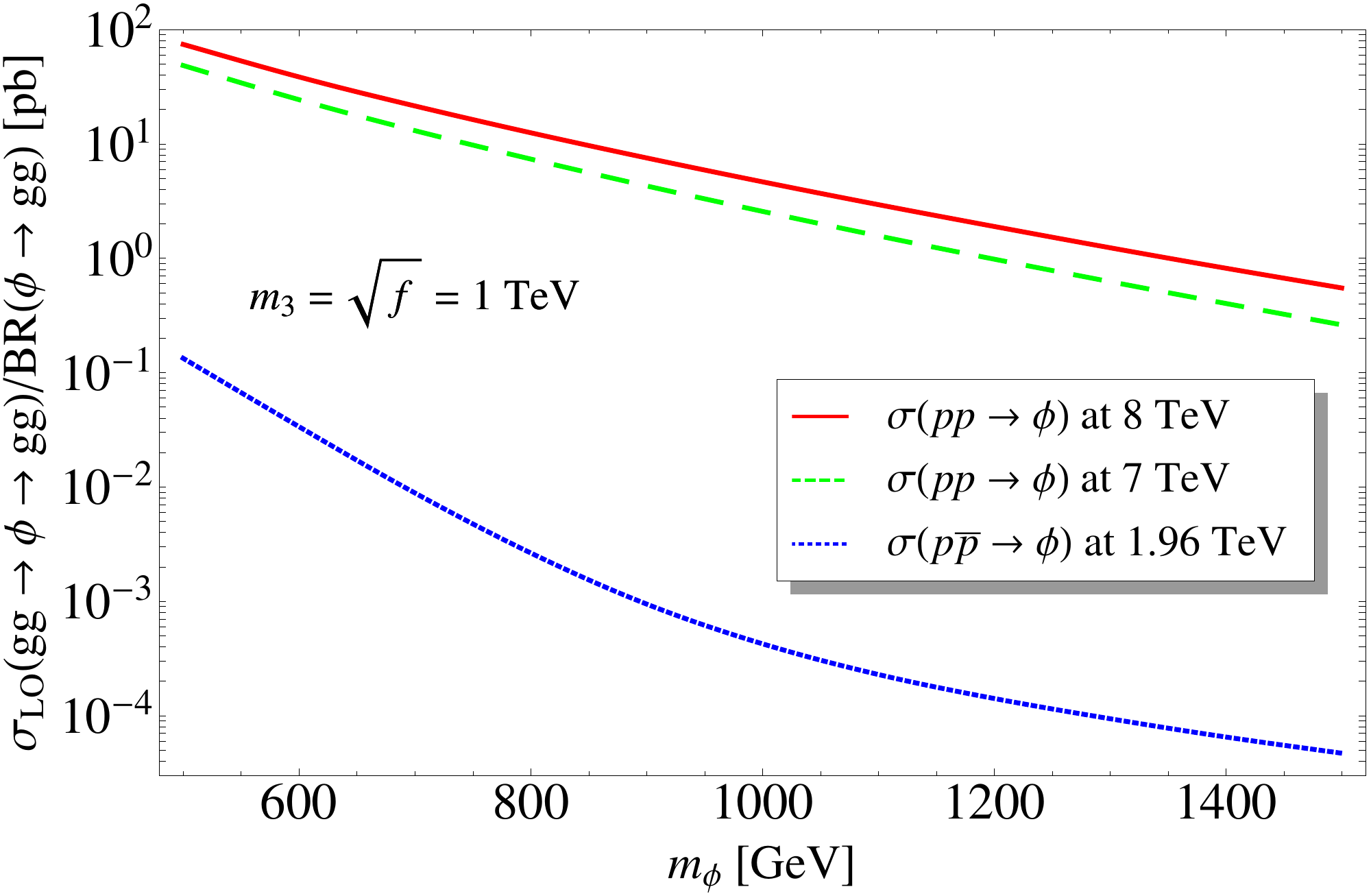}\hspace{1cm}
\includegraphics[scale=0.38]{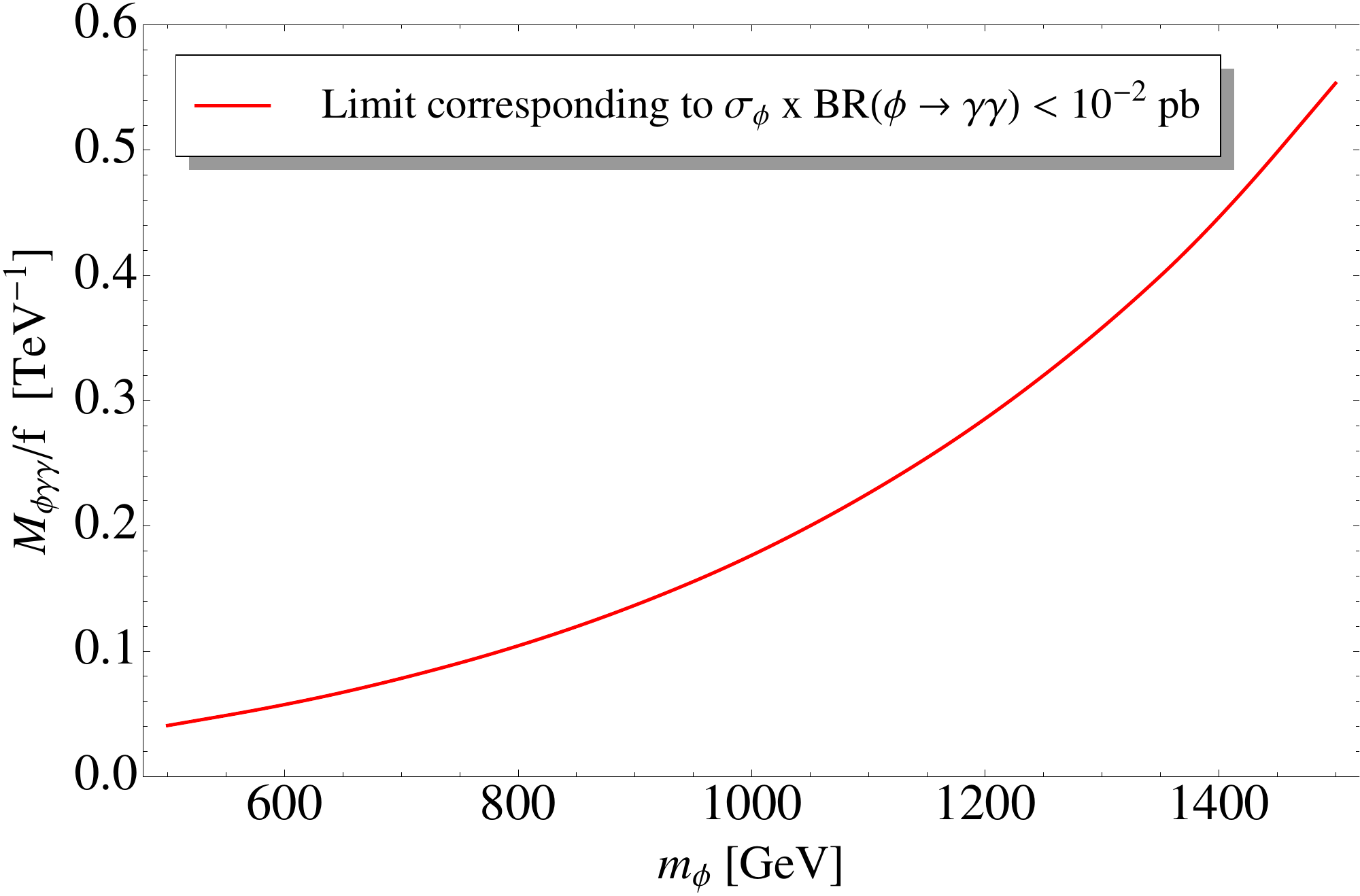}
\end{center}
\caption{ 
\small
Left panel: Sgoldstino production cross sections at LHC@7, LHC@8 and Tevatron; right panel: bound from direct search in the $\gamma\gamma$ final state.
}\label{fig:prodCS}
\end{figure*}
This production cross section is shown in the left panel of Figure \ref{fig:prodCS} as a function of the sgoldstino mass for the choice of parameters $m_{3}=\sqrt{f}=1$ TeV \footnote{To take into account the fact that the NWA breaks down at $m_{\phi}\approx \sqrt{s}$, which is particularly relevant at the Tevatron, we always show the ratio $\sigma\(pp\to \phi\to gg\)/\text{BR}\(\phi\to gg\)$, allowing us to provide unified results independent of the NWA.}. Let us stress that this choice of the value of $\sqrt{f}$ is not related to the parameter space we will consider in the following section, where we discuss how the presence of the sgoldstino can affect the properties of a SM-like Higgs scalar and where only greater values of $\sqrt{f}$ are considered. The reason why we have chosen $\sqrt{f}$ to be equal to the gluino mass in the left panel of Figure \ref{fig:prodCS} is that it allows for simple rescalings of the cross sections for different values of $m_{3}$ and $\sqrt{f}$, according to Eqs.~\eqref{prodCS} and \eqref{widthphigg}. The numerical cross sections have been obtained using the CalcHEP Matrix Element Generator \cite{Pukhov:1999p1261,Pukhov:2004ca,Belyaev:Df-Kj9yc} with the CTEQ6L parton distribution functions.

The sgoldstino mass and couplings are in general not very constrained from experimental bounds. Since the couplings to fermions and to the longitudinal components of the EW gauge bosons are suppressed, bounds from LEP searches and from EW Precision Tests (EWPT) are irrelevant in the region of the parameter space we consider. The only relevant bounds come from direct searches for a resonance in the di-jet, di-photon or di-boson final states. Direct searches for resonances in the di-jet invariant mass spectrum have been performed by the Tevatron (see, e.g., Ref.~\cite{CDFCollaboration:2009p2710}) and the LHC (see, e.g., Refs.~\cite{ATLASCollaboration:2011ww} and \cite{Chatrchyan:2011ns}). These searches do not set any bound on the sgoldstino mass or its coupling to gluons as long as $m_{3}\leq \sqrt{f}$, which is always true in the region of the parameter space we are considering. 

The most constraining bound comes from direct searches in the di-photon final state. Searches for resonances in the di-photon final state are usually performed in the framework of extra dimensions and are restricted to spin-2 resonances\footnote{See, e.g., the recent searches by the ATLAS \cite{ATLASCollaboration:2011ha} and the CMS \cite{CMSCollaboration:2011wk,CMSCollaboration:2011ua} collaborations.}. However, since we do not know the acceptance times efficiency for a scalar particle in the specific analyses, it is not obvious how to use these results in order to set bounds on the sgoldstino. Of course, the simplest thing to do is to assume these quantities to be the same for the scalar and the spin-2 states in order to compare the experimental limit on $\sigma\times \text{BR}$ with the same quantity computed in our model. 

The only analysis setting a bound on $\sigma\times \text{BR}$ (even if a spin-2 particle is considered) is the ATLAS search of Ref.~\cite{ATLASCollaboration:2011ha}. Unfortunately, this bound contains combined limits coming not only from di-photon but also from di-lepton final states. To be conservative, we assume that this bound corresponds only to di-photon events and we get the bound $\sigma_{\phi}\times \text{BR}\(\phi\to\gamma\gamma\)\lesssim 10^{-2}$ pb, which turns out to be almost independent of the sgoldstino mass in the region of interest. In our model, this bound translates into a bound on $M_{\phi\gamma\gamma}/f$, where the $M_{\phi\gamma\gamma}$ is the mass parameter relevant for the sgoldstino coupling to photons and is determined by the bino and wino masses, see Table~\ref{Table1}. This is due to the fact that the production cross section is proportional to $m_{3}^{2}/f^{2}$, while the decay BR into the di-photon channel, being subdominant with respect to the gluon-gluon channel, is approximately proportional to $M_{\phi\gamma\gamma}^{2}/m_{3}^{2}$. Taking into account Eqs.~\eqref{prodCS}, \eqref{widthphigg} and \eqref{widthphigaga} this bound can be written in the following way,
\begin{equation}
\dst \sigma_{\phi}\times BR\(\phi\to\gamma\gamma\)
\approx\dst \ovl{\sigma}_{\phi}\times \f{\text{TeV}^{2}}{f^{2}} \f{M_{\phi\gamma\gamma}^{2}}{8}\lesssim 10^{-2}\,\mathrm{pb}\,,
\end{equation}
where $\ovl{\sigma}_{\phi}$ is the production cross section computed for $m_{3}=\sqrt{f}=1$ TeV. This implies the constraint,
\be
\f{M_{\phi\gamma\gamma}^{2}}{f^{2}}\lesssim \f{8\cdot 10^{-2}\,\mathrm{pb}}{\ovl{\sigma}_{\phi}}\f{1}{\text{TeV}^{2}}\,,
\ee
which is plotted in the right panel of Fig.~\ref{fig:prodCS} as a function of the sgoldstino mass.

\section{Higgs-sgoldstino mixing and Higgs rates}\label{mix}
\begin{figure*}
\begin{center}
\hspace{-1mm}\includegraphics[scale=0.36]{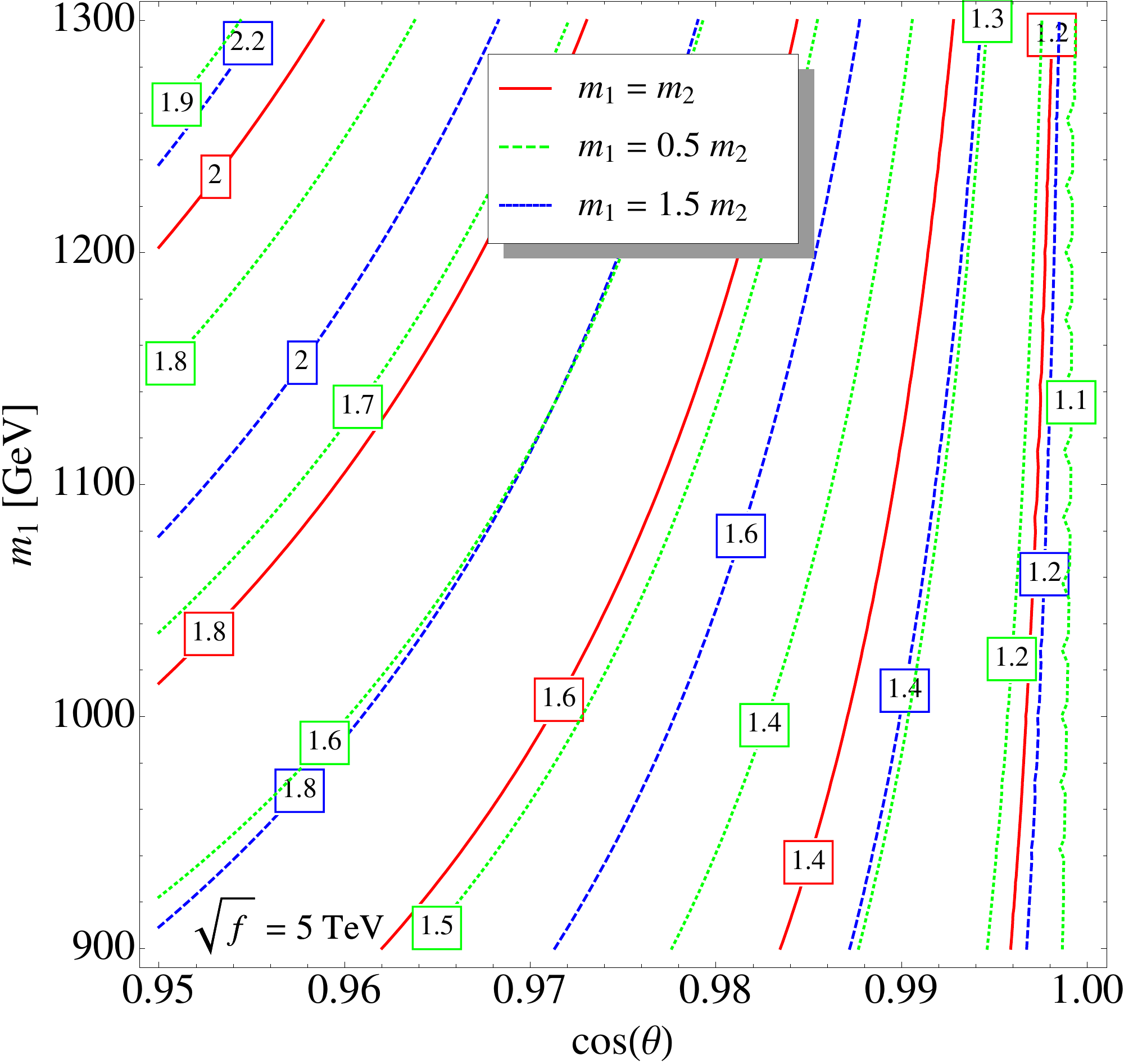}\hspace{1cm}
\includegraphics[scale=0.345]{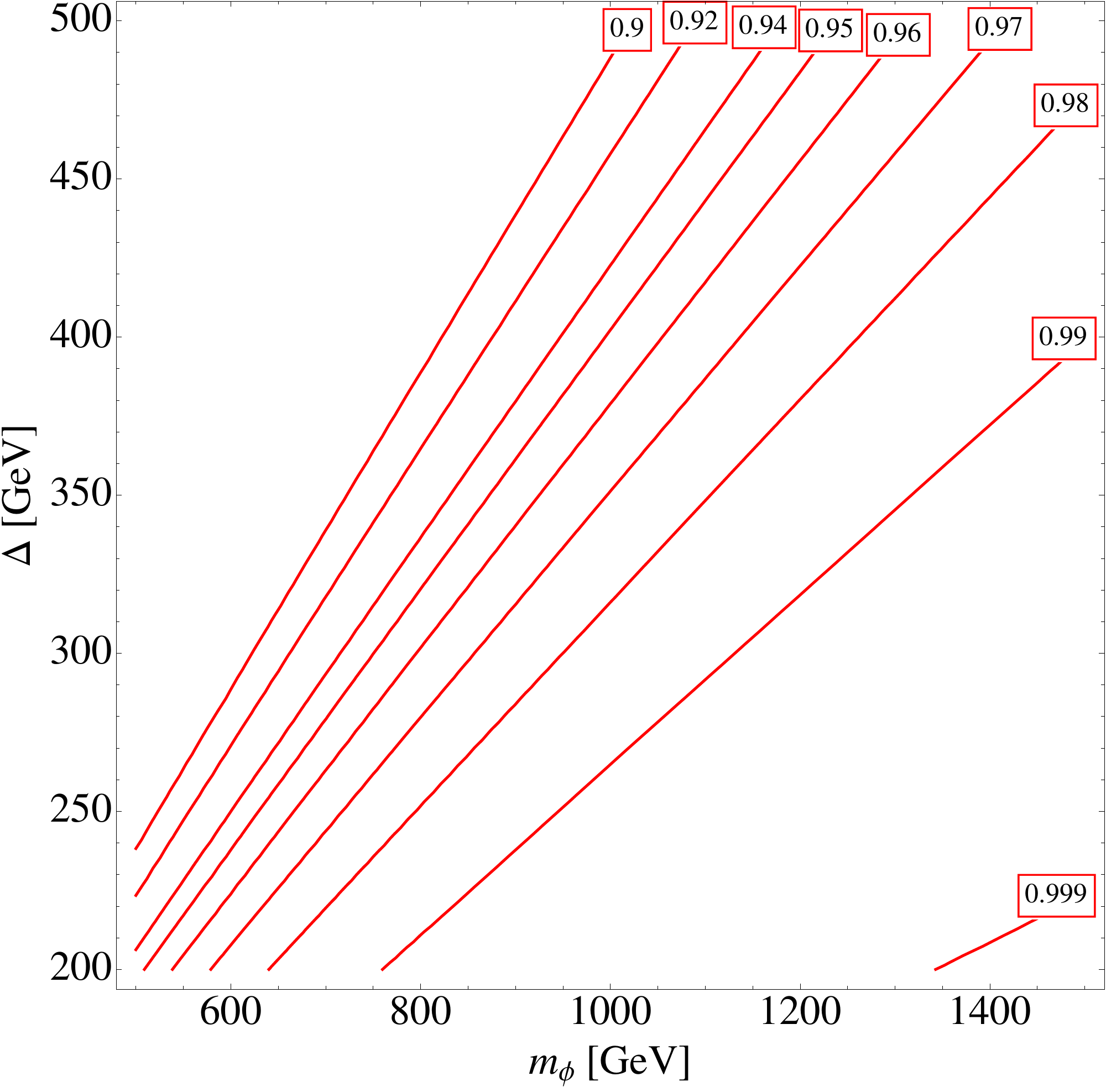}
\end{center}
\caption{ 
\small
Left panel: contours of $\f{\Gamma\(h\to\gamma\gamma\)}{\Gamma\(h\to\gamma\gamma\)_{\text{SM}}}$ in the $(\cos\theta,m_{1})$ plane; right panel: contours of $\cos\theta$ in the $(\Delta,m_{\phi})$ plane.
}\label{fig:enhance}
\end{figure*}
Since the Higgs couplings to photons and gluons are generated only at one loop, even a small mixing with the sgoldstino may significantly affect the Higgs phenomenology at the LHC. In particular, we have seen that the sgoldstino enjoys large tree-level couplings   to the transverse gauge bosons of the form $(m_3/f) \phi\, G_{\mu\nu}G^{\mu\nu}$ and $(M_{\phi\gamma\gamma}/f)\phi\, F_{\mu\nu}F^{\mu\nu}$, whereas the couplings to longitudinal EW bosons and to fermions are in general suppressed.  
Therefore, even a small sgoldstino-Higgs mass mixing $s_\theta$ can generate an $\mathcal{O}(1)$ correction to the $hF_{\mu\nu}F^{\mu\nu}$ vertex which, in turn, gives a sizable correction to the discovery channel $h\rightarrow \gamma\gamma$ if $s_\theta M_{\phi\gamma\gamma}/f=\mathcal{O}(10^{-2})$ TeV$^{-1}$. 
On the other hand, a small mixing will have a negligible effect on the Higgs widths into EW gauge bosons and fermions since their corresponding couplings are already present at tree-level. 

The same sgoldstino-Higgs mixing  contributes also to the gluon fusion production or equivalently to the Higgs partial width to two gluons according to
\be
\f{\delta\Gamma(h\rightarrow gg)}{\Gamma_{\mathrm{SM}}(h\rightarrow gg)}\approx \f{N_c m_3^2}{M^{2}_{\phi\gamma\gamma}}\f{\Gamma_{\mathrm{SM}}(h\rightarrow \gamma\gamma)}{\Gamma_{\mathrm{SM}}(h\rightarrow gg)}\,,
\ee
where $N_c=8$ is the color factor.
Assuming that the rates in EW gauge bosons are within $20-30\%$ of the SM ones we obtain a quite compressed spectrum with at least one of either the bino or wino masses being comparable to the gluino mass. Including  more exotic decays (e.g.~invisible widths)  would allow for a larger production cross-section and relax the compressed spectrum.  As an example, one could arrange for sgoldstino decay into goldstinos which, after mixing, induces an invisible width for the Higgs boson, namely $\Gamma(h\rightarrow GG)=(s_{\theta}^{2}/c_{\theta}^{2})\, \Gamma(\phi \rightarrow GG)$. In the following, for simplicity, we will not consider this possibility any further.

Let us now discuss a simple realization of the scenario discussed above. We consider the decoupling limit of a  SUSY Higgs sector augmented with a sgoldstino. We trade a small mass mixing $\Delta$ for  the mixing angle, \mbox{$s_\theta=\sin\theta\simeq \Delta^2/(m_{h}^2-m_\phi^2)$}, where $m_h$ and $m_\phi$ can be treated as the physical masses as long as $\theta\ll 1$ as is required by having SM-like tree-level vertices. Due to Higgs-sgoldstino mixing, the Higgs couplings to photons and gluons are modified according to
\be
\bry{lll}
\dst \mathcal{L}_{\text{eff}}&=&\dst \left[1+ \frac{s_\theta}{c_\theta} \frac{\pi v M_{\phi\gamma\gamma}}{c_\gamma^{\mathrm{SUSY}} f}\right] c_\theta c_\gamma^{\mathrm{SUSY}}\frac{h}{\pi v}F_{\mu\nu}F^{\mu\nu}\vspace{2mm}\\
&&\dst -\left[1- \frac{s_\theta}{c_\theta}\frac{12 \pi v m_3}{2\sqrt{2} c_g^{\mathrm{SUSY}} f }\right] c_\theta c_{g}^{\mathrm{SUSY}}\frac{h}{12\pi v}G_{\mu\nu}G^{\mu\nu}\,,
\ery
\ee
where the positive coefficients $c^{\mathrm{SUSY}}_{\gamma\,, g}$ parametrize the corresponding vertices in the no-mixing limit.
The left panel in Fig.~\ref{fig:enhance} shows the enhancement in \mbox{$\Gamma(h\rightarrow \gamma\gamma)/\Gamma_{\mathrm{SM}}(h\rightarrow \gamma\gamma)$}  matching $c_\gamma^{\mathrm{SUSY}}$ to the Leading Order SM value $c_\gamma^{\mathrm{SM}}\simeq \alpha$.
The typical mixing angle $s_\theta$ is in the range $0-0.05$ corresponding to a mass mixing $\Delta=200-500$ GeV for $m_\phi$ in the range \mbox{$500-1500$ GeV} (see the right panel in Fig.~\ref{fig:enhance}).
Notice that the effect of the mixing on the Higgs production cross section by gluon-gluon fusion and the Higgs decay to $\gamma\gamma$ can be either constructive or destructive, depending on the sign of the three gaugino masses. In particular, for $m_{3}$ and $M_{\phi\gamma\gamma}$ both positive, it is possible to deplete the Higgs production cross section and reduce the signal rate in all channels except for the $\gamma \gamma$, which, at the same time, can be increased.  

\section{Conclusions}\l{Sec:Conclusion}

In this paper we discussed the phenomenology of a light sgoldstino at the LHC, and how it affects the production and decay rates of a SM-like Higgs boson. 
We showed that a small Higgs-sgoldstino mixing can enhance the Higgs couplings to the gauge bosons because of the sizable tree-level couplings of the sgoldstino to the transverse polarizations of the gauge bosons. 
In particular, we showed that it is possible to achieve an $\mathcal{O}(1)$ enhancement in the di-photon Higgs signal rate by considering a sgoldstino scalar at around the TeV scale, without affecting the other rates  and without conflicting with any other experimental bounds.   
\begin{table}
\caption{\small\label{Table1} The couplings of the CP-even sgoldstino 
to the SM particles and the goldstino in terms of the soft masses.}
\begin{ruledtabular}
\begin{tabular}{c||cc}
\qquad Coupling \,\qquad\qquad & Analytical Expression \\
\hline $\dst M_{\phi g g}$ & $ \dst \f{m_{3}}{2\sqrt{2}} $ \\
\hline $\dst M_{\phi \gamma \gamma}$ & $\dst  \f{m_{1} \cos^{2} \theta_W +m_{2} \sin^{2} \theta_W}{2\sqrt{2}} $ \\
\hline $\dst M_{\phi \gamma Z}$ & $\dst  \f{(m_{2}-m_{1}) \sin \theta_W \cos \theta_W}{\sqrt{2}} $ \\
\hline $\dst M_{\phi Z Z}^{(T)}$ & $\dst  \f{m_{1} \sin^{2} \theta_W +m_{2} \cos^{2} \theta_W}{2\sqrt{2}} $ \\
\hline $\dst M_{\phi Z Z}^{(L)}$ & $\dst  -\f{m_{\phi ZZ}}{\sqrt{2}}  $ \\
\hline $\dst M_{\phi W W}^{(T)}$ & $\dst  \f{m_{2}}{\sqrt{2}}  $ \\
\hline $\dst M_{\phi W W}^{(L)}$ & $\dst  -\f{m_{\phi W W}}{\sqrt{2}}  $ \\
\hline $\dst M_{\phi G G}$ & $\dst  \f{i m_{\phi}}{\sqrt{2\sqrt{2}}}  $ \\
\end{tabular}
\end{ruledtabular}
\end{table}

\section*{Acknowledgments}

We thank Riccardo Argurio, Gabriele Ferretti, Alberto Mariotti, Alberto Romagnoni and Andrea Thamm for discussions. We also thank CERN for the hospitality during the period in which this work was done. The work of C.P. and R.T. was supported by the MICINN under grants FPA2009-07908 and FPA2010-17747.
R.T. was partially supported by the Research Executive Agency (REA) of the European Union under the Grant Agreement number PITN-GA-2010-264564 (LHCPhenoNet).
B.B. is supported by the ERC Advanced Grant no.267985, “Electroweak Symmetry Breaking, Flavour and Dark Matter: One Solution for Three Mysteries” (DaMeSyFla).
\vspace{4mm}
\appendix*


\section{Sgoldstino couplings and widths}\l{appendix}
The most relevant couplings of the CP-even sgoldstino  to SM particles and to the goldstino are described by the following effective Lagrangian \cite{Perazzi:2000ku},
\begin{equation}
\label{ }
\mathcal{L}_{\mathrm{eff}}=\mathcal{L}_{\phi gg}+\mathcal{L}_{\phi \gamma\gamma}+\mathcal{L}_{\phi \gamma Z}+\mathcal{L}_{\phi  Z Z}+\mathcal{L}_{\phi  WW}+\mathcal{L}_{\phi GG}\,,
\end{equation}
where
\bes
\begin{align}
& \mathcal{L}_{\phi gg}=\f{M_{\phi g g}}{f}\,\phi\, G^{a\,\mu\nu}G^{a}_{\mu\nu} \,,
\\
& \mathcal{L}_{\phi \gamma\gamma}=\f{M_{\phi \gamma\gamma}}{f}\, \phi\, F^{\mu\nu}F_{\mu\nu}\,, 
\\
& \mathcal{L}_{\phi \gamma Z} = \f{M_{\phi \gamma Z}}{f}\,\phi\, F^{\mu\nu}Z_{\mu\nu}\,, 
\\
& \mathcal{L}_{\phi  Z Z} = \f{M_{\phi Z Z}^{(T)}}{f}\,\phi\, Z^{\mu\nu}Z_{\mu\nu}\ 
+\f{M_{\phi Z Z}^{(L)}m_{Z}^{2}}{f}\, \phi Z^{\mu} Z_{\mu}\,,
\\
& \hspace{-1mm}\bry{lll} \dst \mathcal{L}_{\phi  WW}&=&\dst \f{M_{\phi W W}^{(T)}}{f}\,\phi\, W^{+\,\mu\nu}W^{-}_{\mu\nu}\\ 
&&\dst +\f{M_{\phi W W}^{(L)}m_{W}^{2}}{f}\,\phi\, W^{+\,\mu}W^{-}_{\mu}\,,\ery
\\
& \mathcal{L}_{\phi  GG} = \f{M_{\phi G G}^{2}}{f}\,\phi\, G\,G 
\end{align}
\ees
where all the mass parameters $M_{i}$ are given by combinations of soft masses according to Table \ref{Table1}.  

\noindent The two body decay widths of the CP-even sgoldstino-like scalar are given by 
\begin{widetext}
\vspace{-6mm}\bes
\begin{align}
& \bry{lll} \dst \Gamma(\phi \to gg) = \f{2M_{\phi g g}^{2} m^{3}_{\phi}}{\pi f^{2}}\,,\l{widthphigg} \ery  \\
& \bry{lll} \dst  \Gamma(\phi \to \gamma\gamma) = \f{ M_{\phi \gamma \gamma}^{2} m^{3}_{\phi}}{4 \pi f^{2}}\,,\l{widthphigaga} \ery\\
& \bry{lll} \dst  \Gamma(\phi \to \gamma Z) = \f{ M_{\phi \gamma Z}^{2} m^{3}_{\phi}}{8\pi f^{2}} \left(1-\f{m_{Z}^{2}}{m_\phi^{2}} \right)^{3}\,,\ery\\
& \bry{lll} \Gamma(\phi \to Z Z) &=& \dst \f{m_{Z}^{4}}{8\pi\,m_\phi f^{2}}\Bigg[ 2\,(M_{\phi Z Z}^{(T)})^{2}  \left(6-\f{4m_\phi^{2}}{m_{Z}^{2}}+\f{m_\phi^4}{m_{Z}^{4}}\right)+ \,12 \, M_{\phi Z Z}^{(T)} \,M_{\phi Z Z}^{(L)} \left(1- \f{m_\phi^{2}}{2m_{Z}^{2}} \right) \\
&& \dst + \, (M_{\phi Z Z}^{(L)})^{2} \left( 3-\f{m_\phi^{2}}{m_{Z}^{2}}+\f{m_\phi^4}{4m_{Z}^4} \right)\Bigg] 
\left( 1-\f{4m_{Z}^{2}}{m_{\phi}^{2}}\right)^{1/2}\,, \ery \\
& \bry{lll} \Gamma(\phi \to WW) &=&  \dst \f{m_{W}^{4}}{8\pi\,m_{\phi} f^{2}}\Bigg[ (M_{\phi W W}^{(T)})^{2}  \left(6-4\f{m_\phi^{2}}{m_{W}^{2}}+\f{m_{\phi}^{4}}{m_{W}^{4}}\right)+12 M_{\phi W W}^{(T)} M_{\phi W W}^{(L)} \left(1- \f{m_\phi^{2}}{2m_{W}^{2}} \right)  \\
&& \dst +\,2 (M_{\phi W W}^{(L)})^{2} \left( 3-\f{m_\phi^{2}}{m_{W}^{2}}+\f{m_\phi^4}{4m_{W}^4}\right)\Bigg]  \left( 1-\f{4m_{W}^{2}}{m_{\phi}^{2}}\right)^{1/2}\,, \ery\\
& \bry{lll} \dst \Gamma(\phi \to GG) = \f{M_{\phi G G}^{4} m_{\phi}}{4\pi f^{2}}\,.\ery
\end{align}
\ees
\end{widetext}

\bibliographystyle{JHEP}
\bibliography{paper}{}

\end{document}